\documentclass[utf8]{frontiersSCNS} 

\usepackage{url,hyperref,lineno,microtype,subcaption}
\usepackage[onehalfspacing]{setspace}
\newcommand{\R}{\ensuremath{\mathbb{R}}}

\def\firstAuthorLast{Bertoni {et~al.}} 
\def\Authors{Federico Bertoni\,$^{1,3,4,5}$, Noemi Montobbio\,$^{2}$,  Alessandro Sarti$^{4,6}$ and  Giovanna Citti\,$^{3,4,6,*}$ }
\def\Address{$^{1}$Sorbonne Universit\'e, Paris, France \\
$^{2}$ Neural Computation Laboratory,  Center for Human Technologies,  Istituto Italiano di Tecnologia,  Genova,  Italy \\
$^{3}$Dipartimento di Matematica, Universit\`{a} di Bologna, Italy \\
$^{4}$CAMS, CNRS - EHESS, Paris, France \\
$^5$This project has received funding from the European Union's Horizon 2020 research and innovation program under the Marie Sk\l odowska-Curie grant agreement No 754362. \includegraphics[height=1cm]{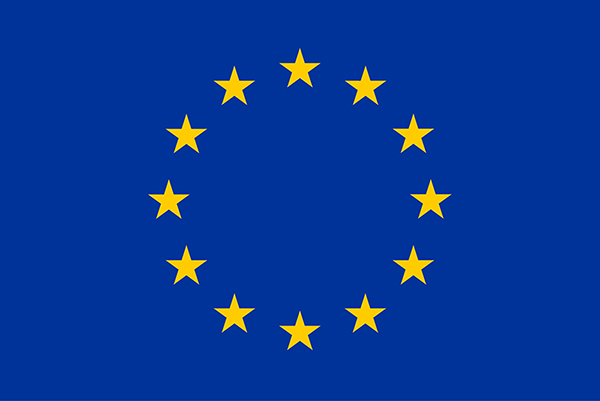} \\
$^{6}$This project has received funding from GHAIA - Marie Sk\l odowska-Curie grant agreement No 777822.}
\def\corrAuthor{ }

\def\corrEmail{ }

\begin{document}
\onecolumn
\firstpage{1}

\title[Emergence of Lie symmetries in functional architectures learned by CNNs]{Emergence of Lie symmetries in functional architectures learned by CNNs } 

\author[\firstAuthorLast ]{\Authors} 
\address{} 
\correspondance{} 

\extraAuth{}

\maketitle

In this paper we study the spontaneous development of symmetries in the early layers of a Convolutional Neural Network (CNN) during learning on natural images. Our architecture is built in such a way to mimic the early stages of biological visual systems. In particular, it contains a pre-filtering step $\ell^0$ defined in analogy with the Lateral Geniculate Nucleus (LGN). Moreover, the first convolutional layer is equipped with lateral connections defined as a propagation driven by a learned connectivity kernel, in analogy with the horizontal connectivity of the primary visual cortex (V1). The layer $\ell^0$ shows a rotational symmetric pattern well approximated by a Laplacian of Gaussian (LoG), which is a well-known model of the receptive profiles of LGN cells. 
The convolutional filters in the first layer can be approximated by Gabor functions, in agreement with well-established models for the profiles of simple cells in V1. We study the learned lateral connectivity kernel of this layer, showing the emergence of orientation selectivity w.r.t. the learned filters. We also examine the association fields induced by the learned kernel, and show qualitative and quantitative comparisons with known group-based models of V1 horizontal connectivity.
These geometric properties arise spontaneously during the training of the CNN architecture, analogously to the emergence of symmetries in visual systems thanks to brain plasticity driven by external stimuli.

\section{Introduction } \label{Sym_Introduction}             

The geometry of the visual system has been widely studied over years, starting from the first celebrated descriptions given by D. H. Hubel and T. N. Wiesel \citep{HW,hubel} and advancing with a number of more recent geometrical models of the early stages of the visual pathway, describing the functional architectures in terms of group invariances \citep{hoffman,petitot,cs06}. Some works have also focused on reproducing processing mechanisms taking place in the visual system using these models -- e.g. detection of perceptual units in \citet{perceptual}, image completion in \citet{Sanguinetti2008}.

On the other hand, relations between Convolutional Neural Networks (CNNs) and the human visual system have been proposed and studied, in order to make CNNs even more efficient in specific tasks \citep[see e.g.][]{Poggio2007}. For instance, in \cite{Yamins2015} and \cite{Yamins2016} the authors were able to study several cortical layers by focusing on the encoding and decoding ability of the visual system, whereas in \citet{Poggio1995,Anselmi2014,Poggio2016} the authors studied some invariance properties of CNNs. A parallel between horizontal connectivity in the visual cortex and lateral connections in neural networks has also been proposed in some works \citep[see e.g.][]{liang,spoerer,Sherstinsky}. Recently, other biologically-inspired modifications of the classical CNN architectures have been introduced \citep[][]{Montobbio,Bertoni}.
 
 In this paper, we combine the viewpoints of these strands of research by studying the emergence of geometrical symmetries in the early layers of a biologically inspired CNN architecture. We will focus on drawing a parallel between the patterns learned from natural images by specific computational blocks of the network, and the symmetries arising in the functional architecture of the Lateral Geniculate Nucleus (LGN) and the primary visual cortex (V1).

\section{Group symmetries in the early visual pathway} \label{Sym_Visual_System}

Over the years, the functional architecture of the early visual pathway has often been modelled in terms of geometric invariances arising in its organization, e.g. in the spatial arrangement of cell tuning across retinal locations, or in the local configuration of single neuron selectivity.
Certain classes of visual cells are shown to act, to a first approximation, as linear filters on an optic signal: the response of one such cell to a visual stimulus $I$, defined as a function on the retina, is given by the integral of $I$ against a function $\psi$, called the \emph{receptive profile} (RP) of the neuron:
\begin{equation}\label{linfilt}
z := \int I(x,y)\psi(x,y)dxdy.
\end{equation}
This is the case for cells in the Lateral Geniculate Nucleus (LGN) and for simple cells in the primary visual cortex (V1) \citep[see e.g. ][]{petitot,cs06}. Both types of cells are characterized by locally supported RPs, i.e. they only react to stimuli situated in a specific retinal region. Each localized area of the retina is known to be associated with a bank of similarly tuned cells \citep[see e.g.][]{HubelWiesel,perceptual}, yielding an approximate invariance of their RPs under translations.

The set of RPs of cells is typically represented by a bank of linear filters $\{\psi_p\}_{p \in \mathcal{G}} \subseteq L^2(\mathbb{R}^2)$ \citep[for some references see e.g.][]{petitot, cs06,symplectic}. The feature space $\mathcal{G}$ is typically specified as a group of transformations of the plane $\{T_p, \;p \in \mathcal{G}\}$ under which the whole filter bank is invariant: each profile $\psi_p$ can be obtained from any other profile $\psi_q$ through the transformation $T_{p-q}$. The group $\mathcal{G}$ often has the product form $\mathbb{R}^2\times\mathcal{F}$, where the parameters $(x_0,y_0) \in\mathbb{R}^2$ determine the retinal location where each RP is centered, while $f\in\mathcal{F}$ encodes the selectivity of the neurons to other local features of the image, such as orientation and scale. 
 
\subsection{Rotational symmetry in the LGN}

A crucial elaboration step for human contrast perception is represented by the processing of retinal inputs via the radially symmetric families of cells present in the LGN \citep{HubelWiesel,hubel}. The RPs of such cells can be approximated by a Laplacian of Gaussian (LoG):
\begin{equation}
\label{LoG}
\psi_{LoG}(x,y) = - \frac{1}{\pi \sigma^4} \left[ 1 - \frac{x^2 + y^2}{2\sigma^2} \right] e^{-\frac{x^2 + y^2}{2\sigma^2}},
\end{equation}
where $\sigma$ denotes the standard deviation of the Gaussian function \citep[see e.g.][]{petitot}.

 \subsection{Roto-translation symmetries in V1 and the lateral connectivity}\label{v1}
 
The invariances in the functional architecture of V1 have been described through a variety of mathematical models. The sharp orientation tuning of simple cells is the starting point of most descriptions. This selectivity is not only found in the response of each neuron to feedforward inputs, but is also reflected in the \emph{horizontal} connections taking place between neurons of V1. These connections show facilitatory influences for cells that are similarly oriented; moreover, the connections departing from each neuron spread anisotropically, concentrating along the axis of its preferred orientation \cite[see e.g.][]{bosking}.

 An established model for V1 simple cells is represented by a bank of Gabor filters $\{\psi_{x_0,y_0,\theta,\sigma}\}_{(x_0,y_0,\theta,\sigma) \in \mathbb{R}^2\times S^1\times\mathbb{R}^2_+}$ built by translations $T_{(x_0,y_0)}$, rotations $R_\theta$ and dilations $D_{\sigma_x, \sigma_y}$ of the filter
 \begin{equation}
 \label{eq:Gabor}
     \psi_{0,0,0,1}(x,y) = A e^{-  \frac{x^2 + y^2}{2}  }   \cos (2 \pi f x + \phi),
 \end{equation}
 where $A$ is the amplitude and $f$ is the frequency of the filter and $\phi$ is the phase which indicates if the Gabor filter is even or odd. See e.g. \citet{daug} and \citet{lee}.
 
 The evolution in time $t\mapsto a(p,t)$ of the activity of the neural population at $p \in \mathbb{R}^2\times \mathcal{F}$ is assumed in \citet{bresscow03} to satisfy a Wilson-Cowan equation \citep{wilcow}:
 \begin{equation}\label{meanfield}
  \partial_t  a(p,t) = -\alpha \: a(p,t) \\
  +\: s\left(\int K(p,p')a(p',t)dp' + z(p,t)\right).
 \end{equation}
 Here, $s$ is a nonlinear activation function; $\alpha$ is a decay rate; $z$ is the feedforward input corresponding to the response of the simple cells in presence of a visual stimulus, as in Eq. (\ref{linfilt}); and the kernel $K$ weights the strength of horizontal connections between $p$ and $p'$. The form of this connectivity kernel has been investigated in a number of studies employing differential geometry tools. A breakthrough idea in this direction has been that of viewing the feature space as a fiber bundle with basis $\mathbb{R}^2$ and fiber $\mathcal{F}$. This approach first appeared in the works of \citet{koenderink} and \citet{hoffman}. It was then further developed by \citet{petitond} and \citet{cs06}. In the latter work, the model is specified as a sub-Riemannian structure on the Lie group $\mathbb{R}^2 \times S^1$ by requiring the invariance under roto-translations. Other works extended this approach by inserting further variables such as scale, curvature, velocity \cite[see e.g.][]{symplectic,abbfav,bccs}.
 
\begin{figure}
\begin{center}
\includegraphics[width=.8\textwidth]{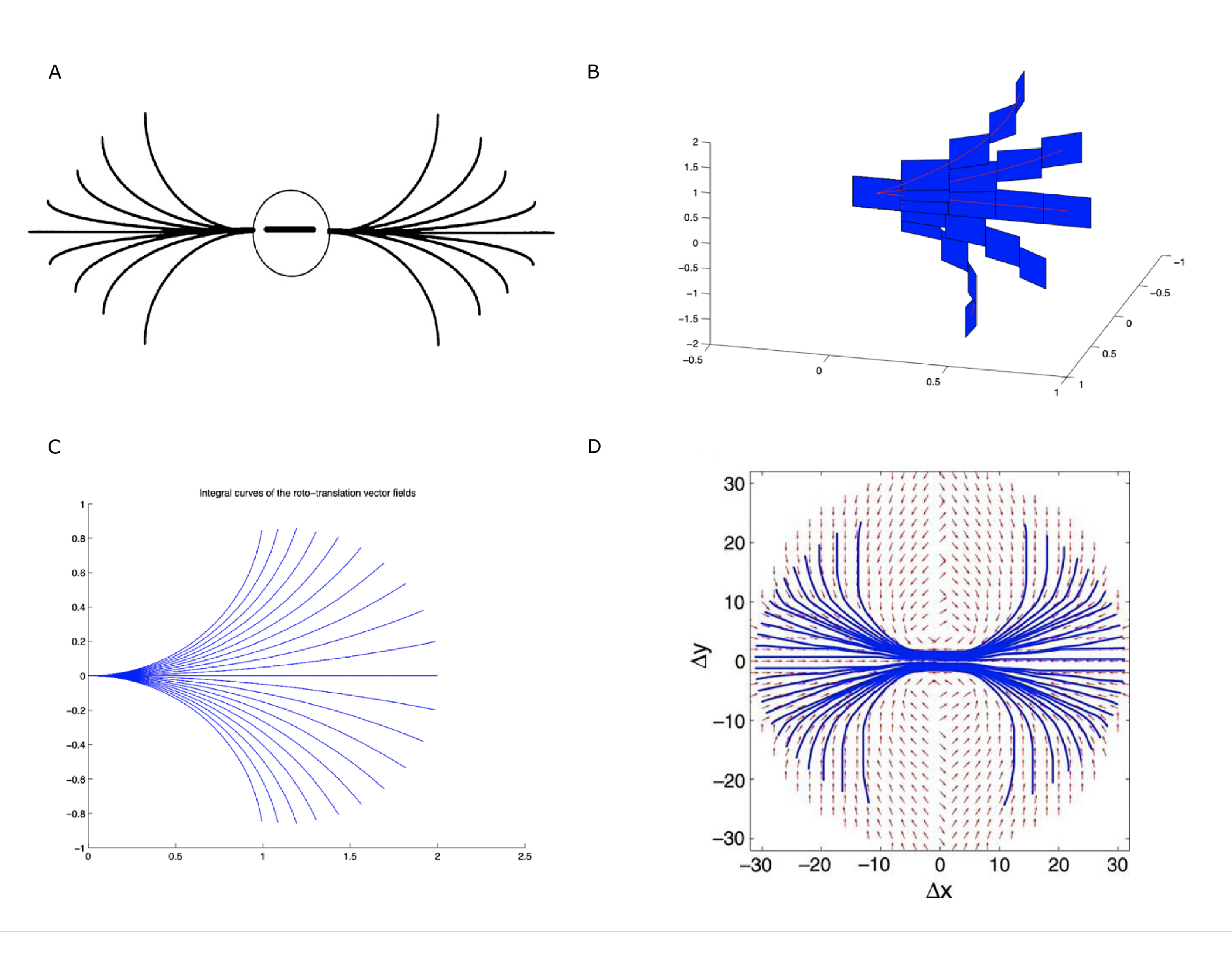}
\end{center}
\caption{\textbf{(A)} Association fields from the experiment of Field, Hayes and Hess \citep{field}. \textbf{(B)} 3D representation of the association field with contact planes of integral curves of the fields (\ref{X1X2}) with varying values of the parameter $k$, from \citep{cs06}. \textbf{(C)} Integral curves of the fields (\ref{X1X2}) with varying $k$, from \citep{cs06}. \textbf{(D)} The vector field of unitary vectors oriented with the maximal edge co-occurrence probability (red) with superimposed its integral curves (blue), from \citep{edge-stat}.}\label{fig:Visual_System}
\end{figure}

 The long-range horizontal connections of V1 are believed to constitute the neural implementation of contour completion, i.e. the ability to group local edge items into extended curves. This perceptual phenomenon has been described through \emph{association fields} \citep{field}, characterizing the geometry of the mutual influences between oriented local elements. See Figure \ref{fig:Visual_System}A from the experiment of Field, Heyes and Hess. Association fields have been characterized in \citet{cs06} as families of integral curves of the two vector fields 
 \begin{equation} \label{X1X2}
    \Vec{X_1} = (\cos \theta, \sin \theta, 0), \;\;\;\; \Vec{X_2} = (0,0,1)
\end{equation}
generating the sub-Riemannian structure on $\mathbb{R}^2 \times S^1$. Figure \ref{fig:Visual_System}B  shows the 3-D constant coefficients integral curves of  the vector fields (\ref{X1X2}) and Figure \ref{fig:Visual_System}C their 2-D projection.
 These integral curves are the solution of the following ordinary differential equation:
\begin{equation*}
    \gamma'(t) = X_1 (\gamma(t)) + k X_2(\gamma(t)).
\end{equation*}
The curves starting from  $(0,0,0)$  can be rewritten explicitly in the following way:
\begin{equation}
    \begin{split}
x = \frac{1}{k} sin(kt), \;\;\;\;\;\;\;
y =  \frac{1}{k} (1-cos(kt)), \;\;\;\;\;\;\;
\theta  =  kt.
    \end{split}
    \label{SanguinettiCurves}
\end{equation}
while integral curves starting from a general point $(x_0,y_0,\theta_0)$ can be generated from equations (\ref{SanguinettiCurves}) by translations $T_{(x_0,y_0)}$ and rotations $R_\theta$.

The probability of reaching a point $(x,y,\theta)$ starting from the origin and moving along the stochastic counterpart of these curves can be described as the fundamental solution of a second order differential operator expressed in terms of the vector fields $\Vec{X_1} $, $\Vec{X_2} $. This is why the fundamental solution of the sub-Riemannian heat kernel or Fokker Planck (FP) kernel have been proposed as alternative models of the cortical connectivity. This perspective based on connectivity kernels was further exploited in  \cite{Montobbio2020}: the model of the cortex was rephrased in terms of metric spaces, and the long range connectivity kernel directly expressed in terms of the cells RPs: in this way a strong link was established between the geometry of long range and feedforward cortical connectivity. 
Finally, in \citet{edge-stat} a strong relation between these models of cortical connectivity and statistics of edge co-occurrence in natural images was proved: the FP fundamental solution is indeed a good model also for the natural image statistics. 
In addition, starting from a connectivity kernel parameterized in terms of position and orientation, in \citet{edge-stat} they obtained  the 2-D vector field represented in red in  Figure \ref{fig:Visual_System}D , whose integral curves (depicted in blue) provide an alternative  model of association fields, learned from images.

\section{The underlying structure: a convolutional neural network architecture }\label{Sym_CNN}

In this section we introduce the network model that will constitute the fundamental structure at the basis of all subsequent analyses. The main architecture consists of a typical Convolutional Neural Network (CNN) for image classification \citep[see e.g.][]{Lawrence,Lecun1998}. CNNs were originally designed in analogy with information processing in biological visual systems. In addition to the hierarchical organization typical of deep architectures, translation invariance is enforced in CNNs by local convolutional windows shifting over the spatial domain. This structure was inspired by the localized receptive profiles of neurons in the early visual areas, and by the approximate translation invariance in their tuning. We inserted two main modifications to the standard model. First, we added a pre-filtering step that mimics the behavior of the LGN cells prior to the cortical processing. Second, we equipped the first convolutional layer with lateral connections defined by a diffusion law via a learned transition kernel, in analogy with the horizontal connectivity of V1. We focused on the CIFAR-10 dataset \citep[][]{Krizhevsky}, since it contains natural images with a large statistics of orientations and shapes \citep[see e.g.][]{Ernst}. We expect to find a strict similarity between the connectivity associated to this transition kernel and one observed by \citep{edge-stat} since they are both learned from natural images.

\subsection{LGN in a CNN}

As described in \cite{Bertoni}, since the LGN pre-processes the visual stimulus before it reaches V1, we aim to introduce a ``layer 0'' that mimics this behavior. To this end, a convolutional layer $\ell^0$  composed by a single filter $\Psi^0$ of size $s^0 \times s^0$, a ReLU function and a batch normalization step is added at the beginning of the CNN architecture. If $\ell^0$ behaves similarly to the LGN, then $\Psi^0$ should eventually attain a rotational invariant pattern, approximating the classical Laplacian-of-Gaussian (LoG) model for the RPs of LGN cells. In such a scenario, the rotational symmetry should emerge spontaneously during the learning process induced by the statistics of natural images, in analogy with the plasticity of the brain. In Section \ref{Sym_Lay1} we will display the filter $\Psi^0$ obtained after the training of the network and test its invariance under rotations.

\subsection{Horizontal connectivity of V1 in a CNN}

Although the analogy with biological vision is strong, the feedforward mechanism implemented in CNNs is a simplified one, overlooking many of the processes contributing towards the interpretation of a visual scene. Our aim is to insert a simple mechanism of horizontal propagation defined by entirely learned connectivity kernels, and to analyze the invariance patterns, if any, arising as a result of learning from natural images.

Horizontal connections of convolutional type have been introduced in previous work through a recurrent formula analogous to (\ref{meanfield}), describing an evolution in time \citep{liang,spoerer}. However, the lateral kernels employed in these works are very localized, so that the connectivity kernel applied at each step only captures the connections between neurons very close-by in space. 
Here we propose a two-step version of this model, where the output of the first layer is first mapped to the corresponding feature space through the feedforward stream, yielding an activation pattern $h^1 = ReLU(z^1)$, and then updated through convolution with a connectivity kernel $K^1$ with a wider support. The new output $\tilde{h}^1$ is defined by averaging between this propagated activation $K^1 \ast h^1$ and the original activation $h^1$, so that our update rule reads:
   \begin{equation}\label{update} \tilde{h}^1 = \frac{1}{2}\Big(K^1 \ast h^1 \:+\: h^1 \Big).\end{equation}
The connectivity kernel $K^1$ is a 4-dimensional tensor parameterized by 2-D spatial coordinates $(i,j)$ and by the indices $(f,g)$ corresponding to all pairs of $\ell^1$ filters. For fixed $f$ and $g$, the function 
\begin{equation}\label{fgkernel}
    (i,j) \mapsto K^1(i,j,f,g)
\end{equation} 
represents the strength of connectivity between the filters $\Psi_f$ and $\Psi_g$, where the spatial coordinates indicate the displacement in space between the two filters. The intuitive idea is that $K^1$ behaves like a ``transition kernel'' on the feature space of the first layer, modifying the feedforward output according to the learned connectivity between filters: the activation of a filter encourages the activation of other filters strongly connected with it. 
Also note that the lateral connections take the form of a linear propagation applied on top of the nonlinear activation function of the first layer. As a consequence, an even wider connectivity may be modelled by iterating the process, while still allowing to recover the resulting long-range kernel via ``self-replication'' through convolution against itself:
\begin{equation*}
\frac{1}{2} \Big(K^1 \ast \tilde{h}^1 \:+\: \tilde{h}^1 \Big) 
= \frac{1}{4} \Big(K^1 \ast K^1 \ast h^1 \:+\: 2 K^1 \ast h^1 \:+\: h^1 \Big).
\end{equation*}

\subsection{Description of the architecture and training parameters} \label{CNN_architecture}

In this section we shall give a detailed overview of our CNN architecture, as well as the training scheme employed. 

Since we wanted a gray scale image to be the input of our neural network, we transformed the CIFAR-10 dataset into gray scale images with mean equal to 0.1307 and standard deviation equal to 0.3081. We split the entire dataset composed by 60000 images in a training set, validation set and test set (40000, 10000 and 10000 images respectively).

The architecture is composed by 11 convolutional layers, each one followed by a ReLU function and a batch normalization layer, and by 3 fully-connected layers. For simplicity we omit the ReLU function and the batch normalization layer in the architecture description. A zero padding was applied in order to keep the spatial dimensions unchanged after each convolutional layer. The architecture was defined as follows:

\begin{itemize}
    \item LGN layer: convolutional layer $\ell^0$ containing a single filter $\Psi^0$ of size $11 \times 11$;
    \item $\ell^1$ convolutional layer: 64 filters of size $7 \times 7$;
    \item First layer lateral connections: connectivity kernel $K^1$ of size $13 \times 13 \times 64 \times 64$;
    \item Max pooling of square size 2;
    \item $\ell^2$, $\ell^3$ and $\ell^4$ convolutional layers: $\ell^2$ contains 64 filters of size $5 \times 5 \times 64$, $\ell^3$ and $\ell^4$ contain 64 filters of size $3 \times 3 \times 64$;
    \item Max pooling of square size 2;
    \item $\ell^5$, $\ell^6$ and $\ell^7$ convolutional layers: $\ell^5$ and $\ell^6$ contain 64 filters of size $3 \times 3 \times 64$, whereas $\ell^7$ contain 128 filters of the same size;
    \item Max pooling of square size 2;
    \item $\ell^8$, $\ell^9$ and $\ell^{10}$ convolutional layers: each one contains 128 filters of size $3 \times 3 \times 128$;
    \item Max pooling of square size 2;
    \item 3 fully-connected layers with 1000, 200 and 10 output units respectively.
\end{itemize}

The neural network was trained using the cross entropy loss function, optimizing it via Stochastic Gradient Descent (SGD) with batch size equal to 64, where the $L^2$ regularization parameter $\lambda$ and the momentum were set to 0.001 and 0.9 respectively.
The learning rate was 0.01 and was decreased over epochs by a factor 10 if the performance of the network on the validation set had not increased in the 10 previous epochs. The training of the neural network was performed in 2 steps: we first trained the feedforward part of the architecture, i.e. excluding the lateral connections of $\ell^1$; we then inserted them and trained the connectivity kernel $K^1$ keeping the rest of the weights fixed.

In order to increase the performance avoid overfitting, along with the weight regularization and the batch normalization layers, we applied dropout with dropping probability equal to 0.5 in the feedforward training after convolutional layer $\ell^{10}$, and 0.2 in the second training step after applying the kernel $K^1$. We also used an early stopping approach by terminating the optimization algorithm as soon as the performance on the validation set failed to increase for 50 consecutive epochs. The maximum number of epochs was set to 400.
The network architecture and optimization procedure were coded using Pytorch \citep[][]{Pytorch}.

We want to outline that, since we are interested in the emergence of symmetries, we did not focus on reaching state-of-the-art performances on classifying the CIFAR-10 dataset. However, our architecture reaches fair performances that can be increased by adding further convolutional layers and/or increasing the number of units. We trained several CNNs varying the numbers of layers and features. 

\begin{figure}
\centering
 \includegraphics[width=.5\linewidth]{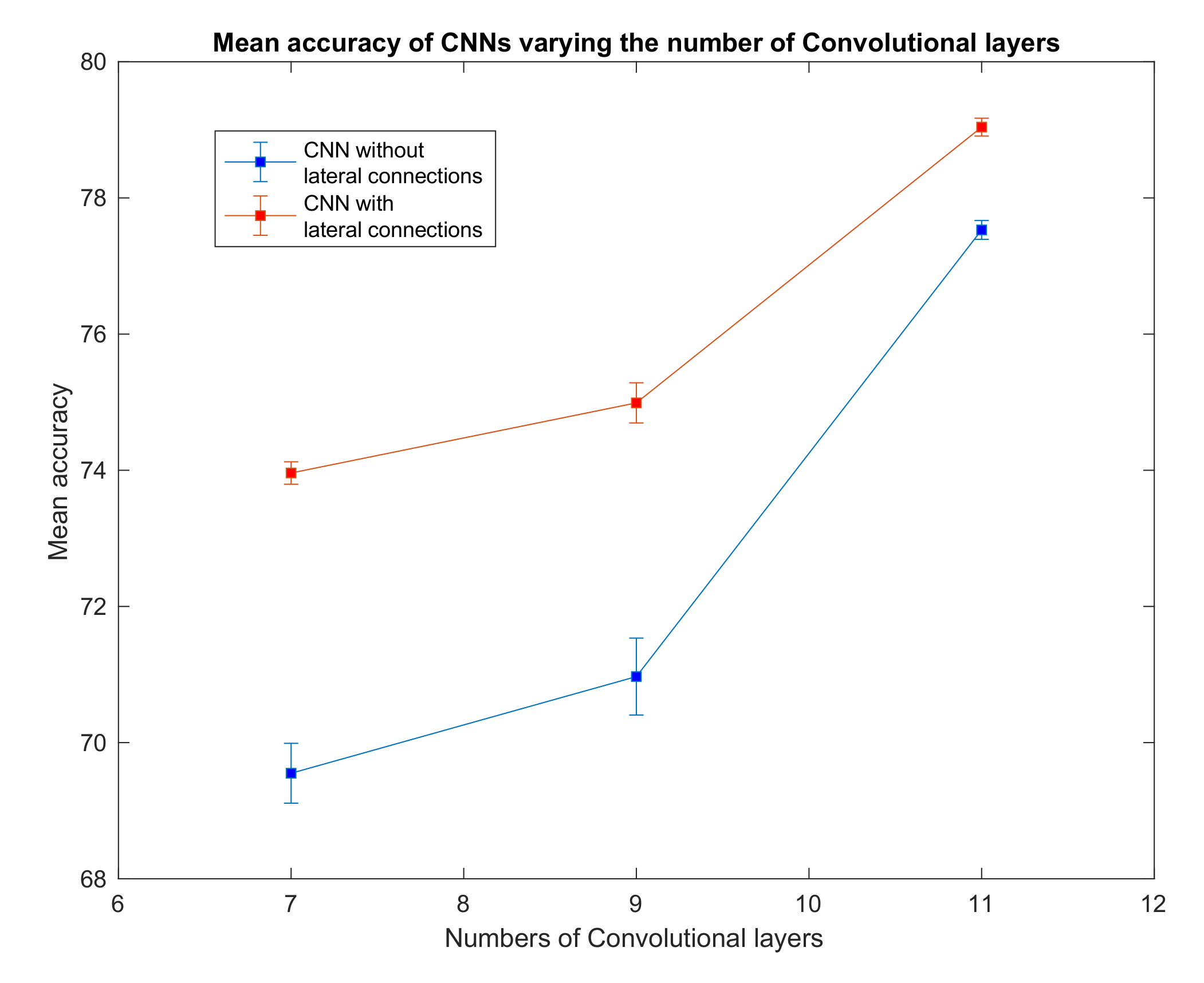}
  \caption{Mean accuracy of CNNs with varying number of convolutional layers, with and without lateral connections in the first layer. Error bars represent the standard error of the mean.}
  \label{Fig:Mean_acc_CNNs}
\end{figure}

Figure \ref{Fig:Mean_acc_CNNs} shows the mean performances on the test set over 20 such models with and without the lateral connectivity. As expected, increasing the number of convolutional layers and adding the lateral connection (and also the number of units) led to better performance. For the analyses described in the following, we chose the architecture that reached the best performances (79.04 \% over 20 models with Horizontal connectivity), i.e. the one with 11 convolutional layers described above. 
However, we stress that all the architectures examined showed a comparable behavior as regards the invariance properties of the convolutional layers $\ell^0$ and $\ell^1$ and the connectivity kernel $K^1$. Therefore, it would be reasonable to expect a similar behavior also when further increasing the complexity of the model.

\section{Emergence of rotational symmetry in the LGN layer} \label{Sym_LGN}

As described in \citep{Bertoni} the introduction of a convolutional layer $\ell^0$ containing a single filter $\Psi^0$ mimics the role of the LGN that pre-filters the input visual stimulus before it reaches the V1 cells.  The authors have shown that a rotational invariant pattern is attained by $\Psi^0$  for a specific architecture, suggesting that it should happen also for deeper and more complex architectures. 

\begin{figure}
\centering
 \includegraphics[width=.47\linewidth]{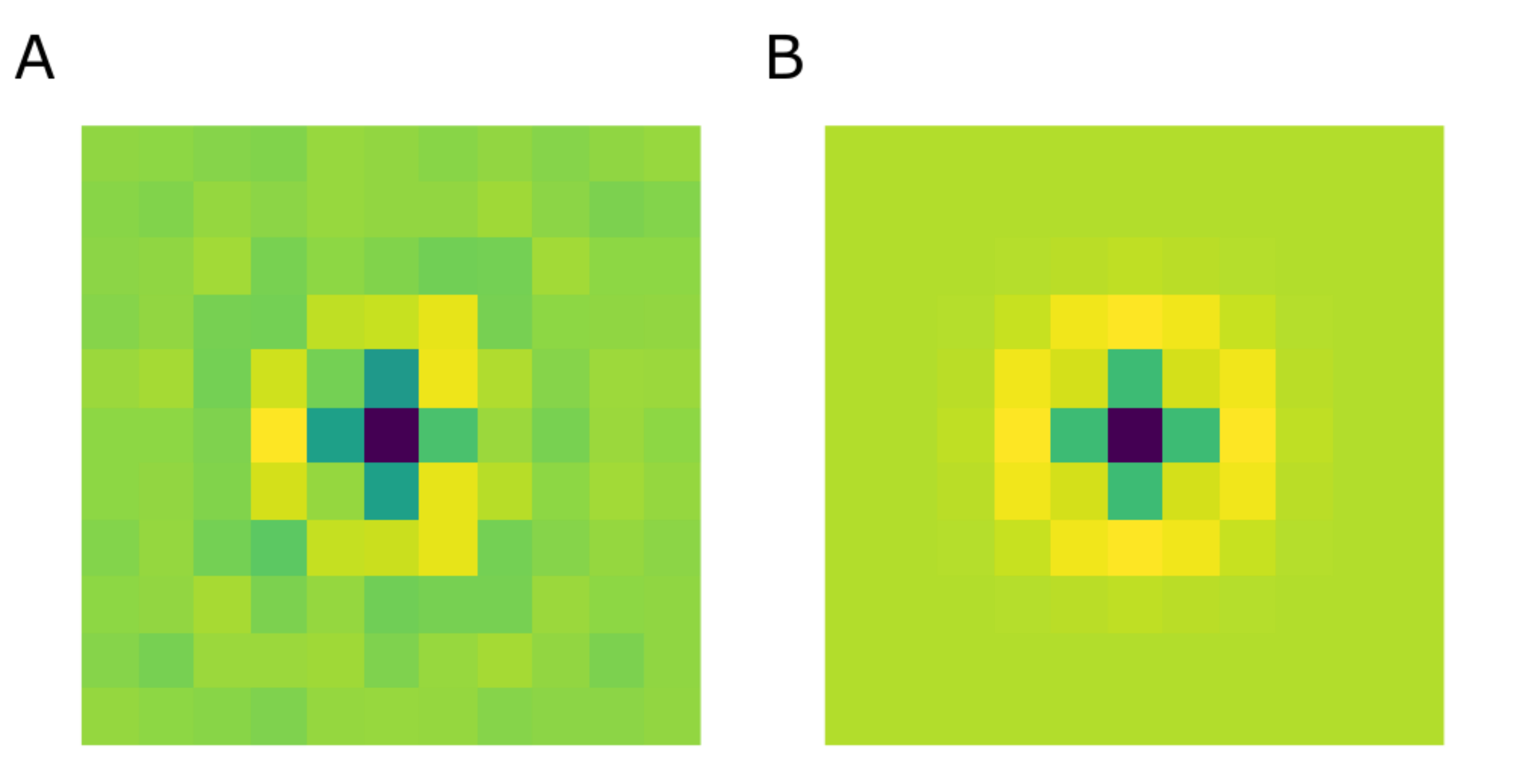}
  \caption{\textbf{(A)} The learned filter $\Psi^0$ of the current architecture. \textbf{(B)} Its approximation as a LoG, with optimal $\sigma = 0.184$, yielding a correlation of 93.67\% with the learned filter.}
  \label{Fig:LoG_approx}

\end{figure}

Figure \ref{Fig:LoG_approx}A shows the filter $\Psi^0$ obtained after the training phase.  As expected it has a radially symmetric pattern and its maximum absolute value is attained in the center.  Thus it can be approximated by the classical LoG model for the RP of an LGN cell by finding the optimal value for the parameter $\sigma$ in Eq. (\ref{LoG}). We used the built-in function \textit{optimize.curve\_fit} from the Python library SciPy.  Figure \ref{Fig:LoG_approx}B shows this approximation with $\sigma = 0.184$.  Applying the built-in function \textit{corrcoef} from the Python library NumPy, it turns out that the two functions have a high correlation of 93.67\%.

These results show that $\Psi^0$ spontaneously evolves into a radially symmetric pattern during the training phase, and more specifically its shape approximates the typical geometry of the RPs of LGN cells.

\section{Emergence of Gabor-like filters in the first layer} \label{Sym_Lay1}

As introduced in Section \ref{v1}, the RPs of V1 simple cells can be modeled as Gabor functions by Eq. (\ref{eq:Gabor}).  Moreover, the first convolutional layer of a CNN architecture usually shows Gabor-like filters \citep[see e.g.][]{Poggio2007}, assuming a role analogous to V1 orientation-selective cells.  In this section, we first approximate the filters of $\ell^1$ as a bank of Gabor filters, in order to obtain a parameterization in terms of position $(x_0,y_0)$ and orientation $\theta$. This will provide a suitable set of coordinates for studying the corresponding lateral kernel in the $\R^2 \times S^1$ domain, see Section \ref{Sym_Kernel}.

\subsection{Approximation of the filters as Gabor functions} \label{subs:Appr_Gabor_L1}

\begin{figure}
\centering
\;\; \includegraphics[width=.8\linewidth]{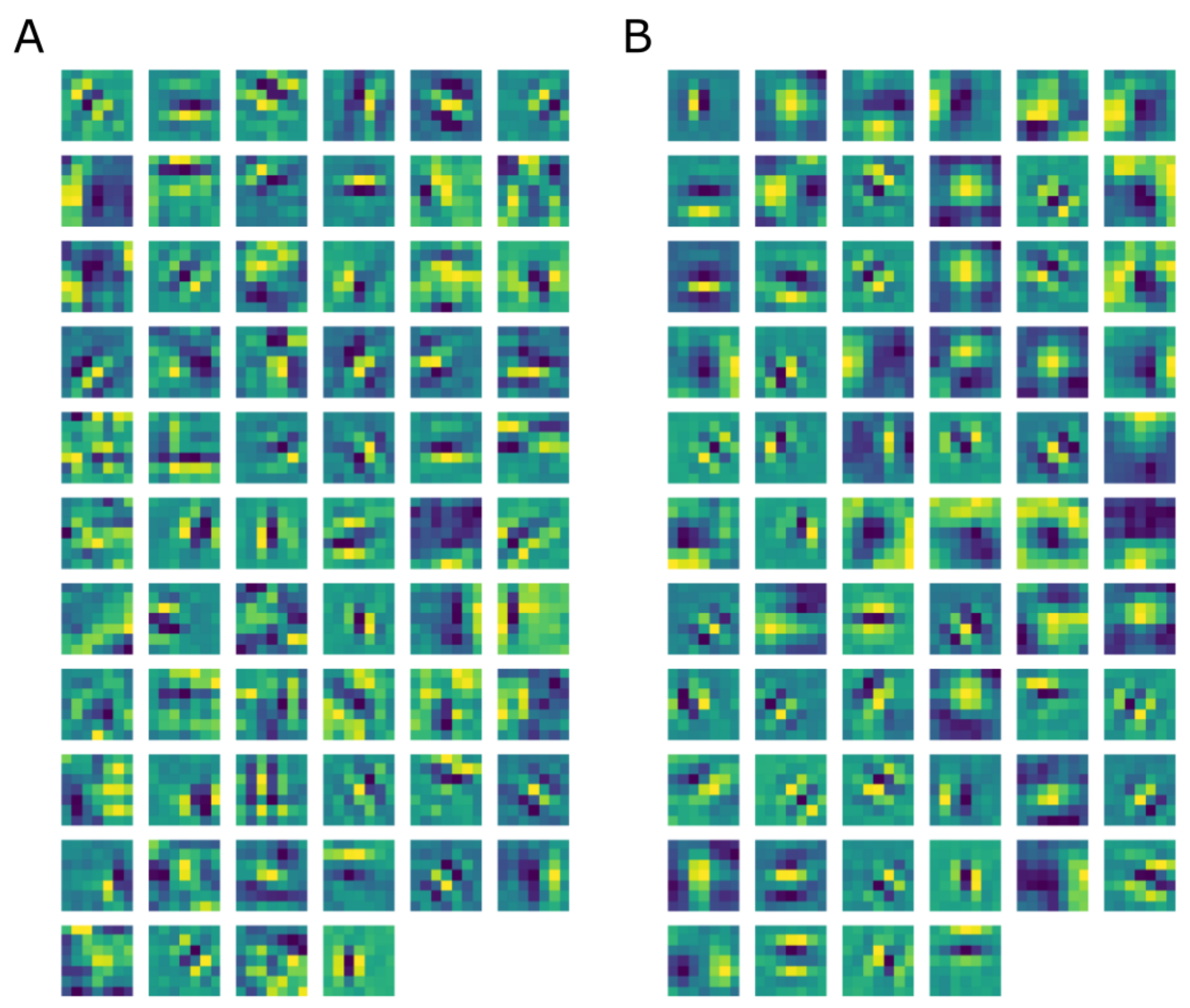}
  \caption{\textbf{(A)} Learned filters of $\ell^1$ of our CNN architecture.
\textbf{(B)} Learned filters of $\ell^1$ of the same CNN architecture, but without $\ell^0$.}
  \label{Fig:L1_NoLGN}
\end{figure}

After the training phase, Gabor-like filters emerge in $\ell^1$ as expected (see Figure \ref{Fig:L1_NoLGN}A). Moreover, thanks to the introduction of the pre-filtering $\ell^0$, our layer $\ell^1$ only contains filters sharply tuned for orientation, with no Gaussian-like filters. Indeed, by removing $\ell^0$, the two types of filters are mixed up in the same layer. See Figure \ref{Fig:L1_NoLGN}B, showing the $\ell^1$ filters of a trained CNN with the same architecture, but without $\ell^0$.

However we can note from Figure \ref{Fig:L1_NoLGN}A that some filters have more complex shapes that are neither Gaussian-, nor Gabor-like; indeed, since we have not introduced other geometric structures on following layers, it is reasonable to observe the emergence of more complex patterns. 
For the subsequent analyses, we considered the 48 filters with highest correlation with the approximating Gabor functions.

The filters in $\ell^1$ were approximated by the Gabor function in Eq. (\ref{eq:Gabor}), where all the parameters were found using the built-in function \textit{optimize.curve\_fit} from the Python library SciPy. The mean correlation between the selected filters and their Gabor approximations obtained using the built-in function \textit{corrcoef} from the Python library NumPy is 92.61\%. 

We then split the filter bank w.r.t. the parity of their approximation, indicated by the parameter $\phi$, that was forced to be between $-\pi$ and $\pi$. Specifically, we labelled a filter as odd if $\frac{\pi}{4} < | \phi | < \frac{3\pi}{4}$, as even if $0<| \phi | < \frac{\pi}{4}$ or $\frac{3\pi}{4} < |\phi| < \pi$.

\begin{figure}
\centering
 \includegraphics[width=\linewidth]{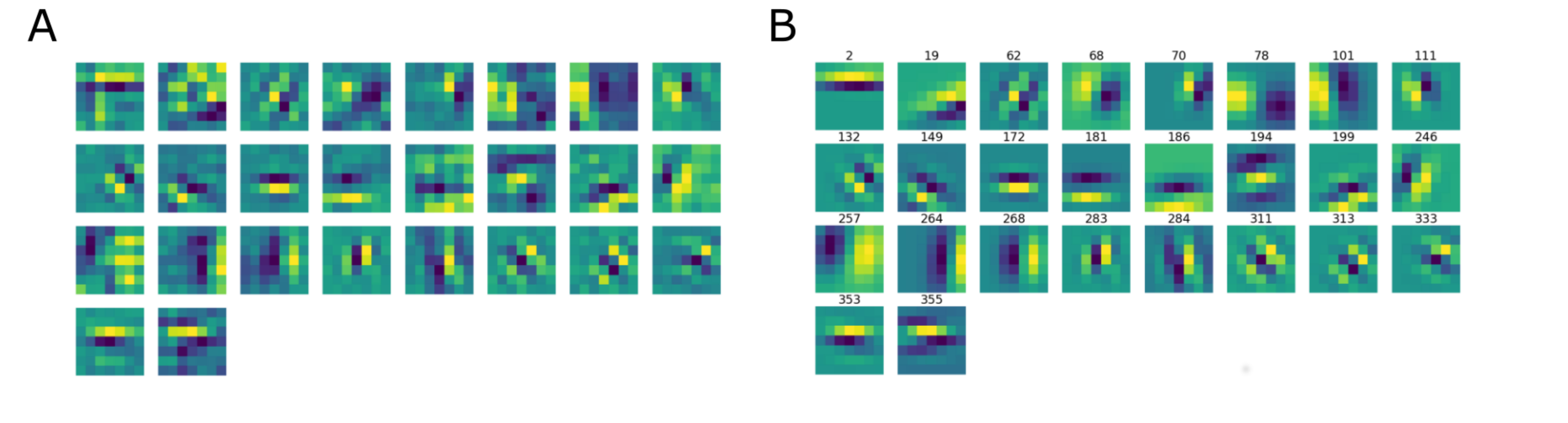}
  \caption{\textbf{(A)}: the learned filters of $\ell^1$ with odd parity, ordered w.r.t. the orientation $\theta$ obtained from the Gabor approximation. \textbf{(B)}: the approximating odd Gabor filters, labelled by their orientation.}
  \label{Fig:Filters_approximation_Odd}

\end{figure}

Figure \ref{Fig:Filters_approximation_Odd} shows the odd filters and their Gabor approximations, rearranged w.r.t. the orientation $\theta$. In this case the orientation $\theta$ spans across the range $[0^\circ, 360^\circ[$, in order to correctly code for filters sharing the same direction but with opposite positive and negative lobes -- and therefore having orientations differing by $180^\circ$. 

\begin{figure}
\centering
 \includegraphics[width=\linewidth]{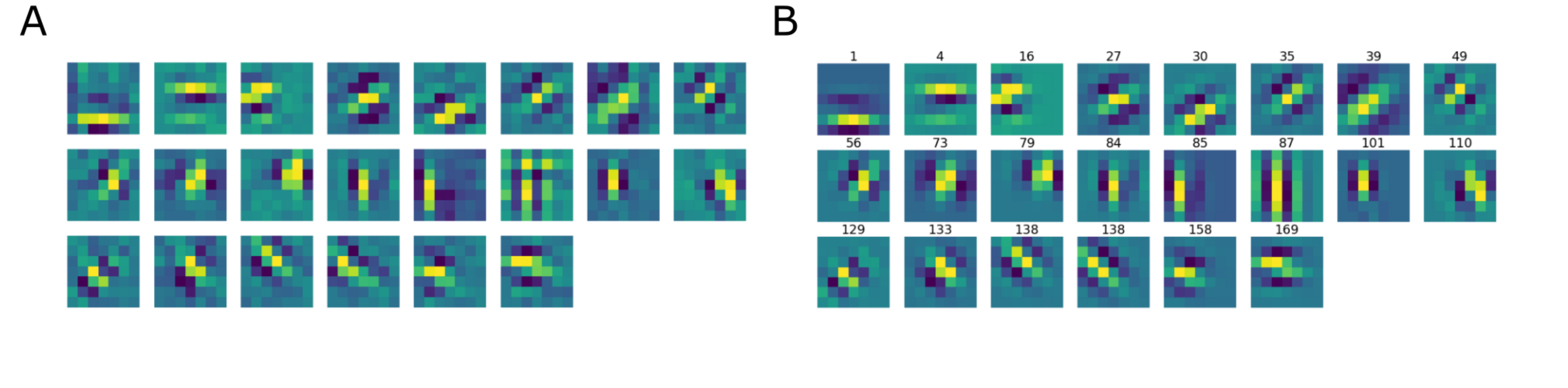}
  \caption{ 
  \textbf{(A)}: the learned filters of $\ell^1$ with even parity, ordered w.r.t. the orientation $\theta$ obtained from the Gabor approximation. \textbf{(B)}: the approximating even Gabor filters, labelled by their orientation.}
  \label{Fig:Filters_approximation_Even}

\end{figure}

Figure \ref{Fig:Filters_approximation_Even} shows the even filters and their Gabor approximations, rearranged w.r.t. the orientation $\theta$. In this case, the orientation $\theta$ spans across the range $[0^\circ, 180^\circ[$. For the sake of visualization, those filters whose central lobe had negative values were multiplied by -1.

In both scenarios, the orientation values are quite evenly sampled, allowing the neural network to detect even small orientation differences. Most of the filters have high frequencies, allowing them to detect thin boundaries, but some low-frequency filters are still present.

\section{Emergence of horizontal connectivity in the transition kernel} \label{Sym_Kernel}           

In this section, we shall examine the learned transition kernel $K^1$, to investigate whether it shows any invariances compatible with the known properties of the lateral connectivity of V1.

\subsection{Re-parameterization of the connectivity kernel using the first layer approximation}

In order to study the selectivity of the connectivity kernel to the properties of $\ell^1$ filters, we first rearranged it based on the set of coordinates in $\mathbb{R}^2\times S^1$ induced by the Gabor approximation of the filters.

We first split the kernel w.r.t. the parity of $\ell^1$ filters, resulting in two separate connectivity kernels for even and odd filters. We then adjusted the spatial coordinates of each kernel using the estimated Gabor filter centers $(x_0, y_0)$. Specifically, for each $f,g$ in $\{1,\ldots,n\}$, we shifted the kernel $K^1(\cdot,\cdot,f,g)$ of Eq. (\ref{fgkernel}) so that a displacement of $(i,j)=(0,0)$ corresponds to the situation where the centers of the filters $\Psi_f$ and $\Psi_g$ coincide. Finally, the original ordering of $f,g$ in $\{1,\ldots,n\}$ has no geometric meaning. However, each filter $\Psi_f$ is now associated with an orientation $\theta_f$ obtained from the Gabor approximation. Therefore, we rearranged the slices of $K^1$ so that the $f$ and $g$ coordinates were ordered by the corresponding orientations $\theta_f$ and $\theta_g$.
By fixing one filter $\Psi_f$, we then obtained a 3-D kernel
\begin{equation}\label{3dkernel}
    (i,j,g) \mapsto K^1(i,j,f,g)
\end{equation}
defined on $\R^2 \times S^1$, describing the connectivity between $\Psi_f$ and all the other $\ell^1$ filters with the same parity properties, each shifted by a set of local displacements $(i,j)\in\{-6,\ldots,6\}\times\{-6,\ldots,6\}$.

\subsection{Association fields induced by the connectivity kernel}
\label{subs:AF}

Starting from the re-parameterized kernel centered around a filter $\Psi_f$ as in Eq. (\ref{3dkernel}), we used the $\theta$-coordinates to construct a 2-D association field as in \cite{edge-stat}. We first defined a 2-D vector field by projecting down the orientation coordinates weighted by the kernel values. Specifically, for each spatial location $(i,j)$, we defined 
\begin{equation}
    V(i,j) := \max_g K^1(i,j,f,g) \:\cdot\:\frac{\sum_{g=1}^n K^1(i,j,f,g) v_g}{\|\sum_{g=1}^n K^1(i,j,f,g) v_g\|},
\end{equation}
where $v_g \in \mathbb{R}^2$ is a unitary vector with orientation $\theta_g$. This yields for each point $(i,j)$ a vector whose orientation is essentially determined by the leading $\theta$ values in the fiber, i.e. the ones were the kernel has the highest values. The norm of the vector is determined by the maximum kernel value over $(i,j)$. Finally, we defined the association field as the integral curves of the so-obtained vector field $V$ starting from points along the trans-axial direction in a neighborhood of $(0,0)$.

\begin{figure}
\centering
\includegraphics[width=1\linewidth]{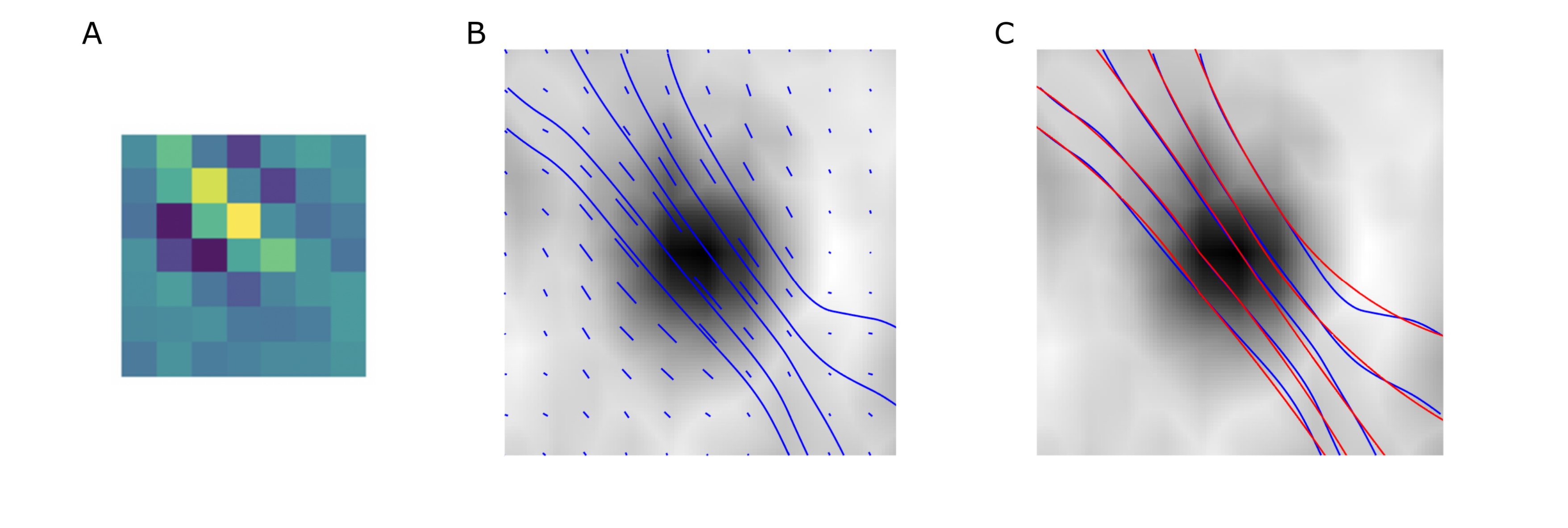}
  \caption{\textbf{(A)} The $7 \times 7$ even filter $\Psi_f$, with orientation $\theta_f=138^\circ$.  
  \textbf{(B)} The vector field $V$ and its integral curves obtained from the kernel of even filters computed around $\Psi_f$, with $\theta_f=138^\circ$. 
\textbf{(C)} The association field of the transition kernel of even filters (blue), and its approximation using the integral curves defined in \eqref{SanguinettiCurves} (red).
For better visualization the kernel has been resized by a factor of 10 using the built-in Matlab function \textit{imresize}.}
  \label{Fig:AF_kernel_Sanguinetti}
\end{figure}

Figure \ref{Fig:AF_kernel_Sanguinetti}B shows the association field obtained from the kernel computed around the even filter $\Psi_f$ in Figure \ref{Fig:AF_kernel_Sanguinetti}A, with orientation $\theta_f=138^\circ$. 
The vector field $V$ was plotted using the Matlab function \textit{quiver}, and the integral curves were computed using the Matlab function \textit{streamline}. The vectors and curves are plotted over a 2-D projection of the kernel obtained as follows. The kernel was first resized by a factor of 10 using the built-in Matlab function \textit{imresize} for better visualization, and then projected down on the spatial coordinates by taking the maximum over $g$. Note that around $(0,0)$ the field is aligned with the orientation of the starting filter $\Psi_f$, while it starts to rotate when it moves away from the center -- consistently with the typical shape of psychophysical association fields, see Section \ref{v1}. 

We also outline the similarity with the integral curves of \cite{edge-stat}, see Figure \ref{fig:Visual_System}D. However, in our case the rotation is less evident since the spatial displacement encoded in the kernel $K^1$ is more localized than the edge co-occurrence kernel constructed in \cite{edge-stat}. 

Figure \ref{Fig:AF_kernel_Sanguinetti}C shows in red the approximation of each integral curve as a circular arc, obtained by fitting the parameter $k$ of Eq. (\ref{SanguinettiCurves}) to minimize the distance between the two curves. The empirical curves induced by the learned connectivity kernel were very close to the theoretical curves, with a mean Euclidean distance of $0.0099$.

\subsection{Comparison of  transition kernel with the solution of heat equation}

\begin{figure}
\centering

   \includegraphics[width=.6\linewidth]{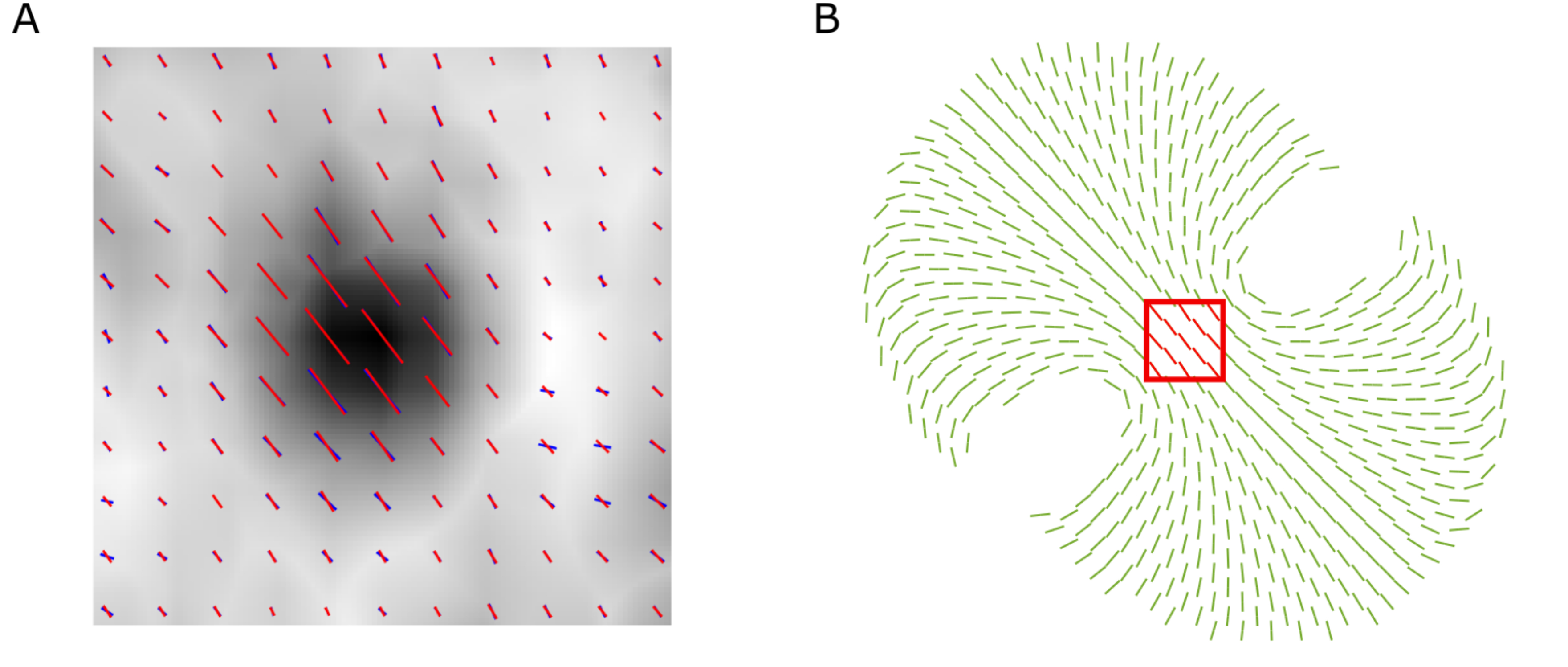}
  \caption{\textbf{(A)} In blue: the vector field of the transition kernel of even filters around $\Psi_f$. In red: the vector field of the best-fitting fundamental solution of the sub-Riemannian heat equation, with $\alpha = 2.74$.
  \textbf{(B)}  In green: the vector field of the best-fitting fundamental solution of the sub-Riemannian heat equation over a wider spatial region. In red: the local vector field fitted to the learned kernel.}
  \label{Fig:Kernel_selection}

\end{figure}

In section 2.2 we recalled the relationship  between connectivity kernel and the fundamental solution of Sub-Riemannian operators. In analogy with this remark, we compare 
 the learned transition kernel $K^1$ with the fundamental solution of the sub-Riemannian heat equation in $\mathbb{R}^2\times S^1$ 
\begin{equation}
\label{eq:Heat}
    \frac{\partial u}{\partial t} = -\alpha \Delta_{X_1,X_2} u = -\alpha \left(X_1^2 u + X_2^2u\right),
\end{equation}
where the sub-Laplacian operator is generated by the vector fields introduced in \cite{cs06} and defined by Eq. (\ref{X1X2}). 
To this end, we fitted the parameter $\alpha$ by comparing the vector field $V$ generated from the kernel $K^1$ with the 2-D vector field obtained with the same procedure from the fundamental solution of (\ref{eq:Heat}). 
More precisely, the optimal parameter $\alpha$ was chosen as the one minimizing the mean angular difference between the two vector fields. We obtained an optimal value $\alpha = 2.74$, with a mean angular difference of $0.0154$. Figure \ref{Fig:Kernel_selection}A shows the vector field $V$ in blue, and the sub-Riemannian vector field with optimal $\alpha$ in red.
As remarked above, the kernel $K^1$ captures the connections over a localized spatial area. Figure \ref{Fig:Kernel_selection}B shows in green the sub-Riemannian vector field over a wider spatial region, with the local area that best fits the vector field $V$ highlighted in red. 

\begin{figure}
\centering
  \includegraphics[width=.8\linewidth]{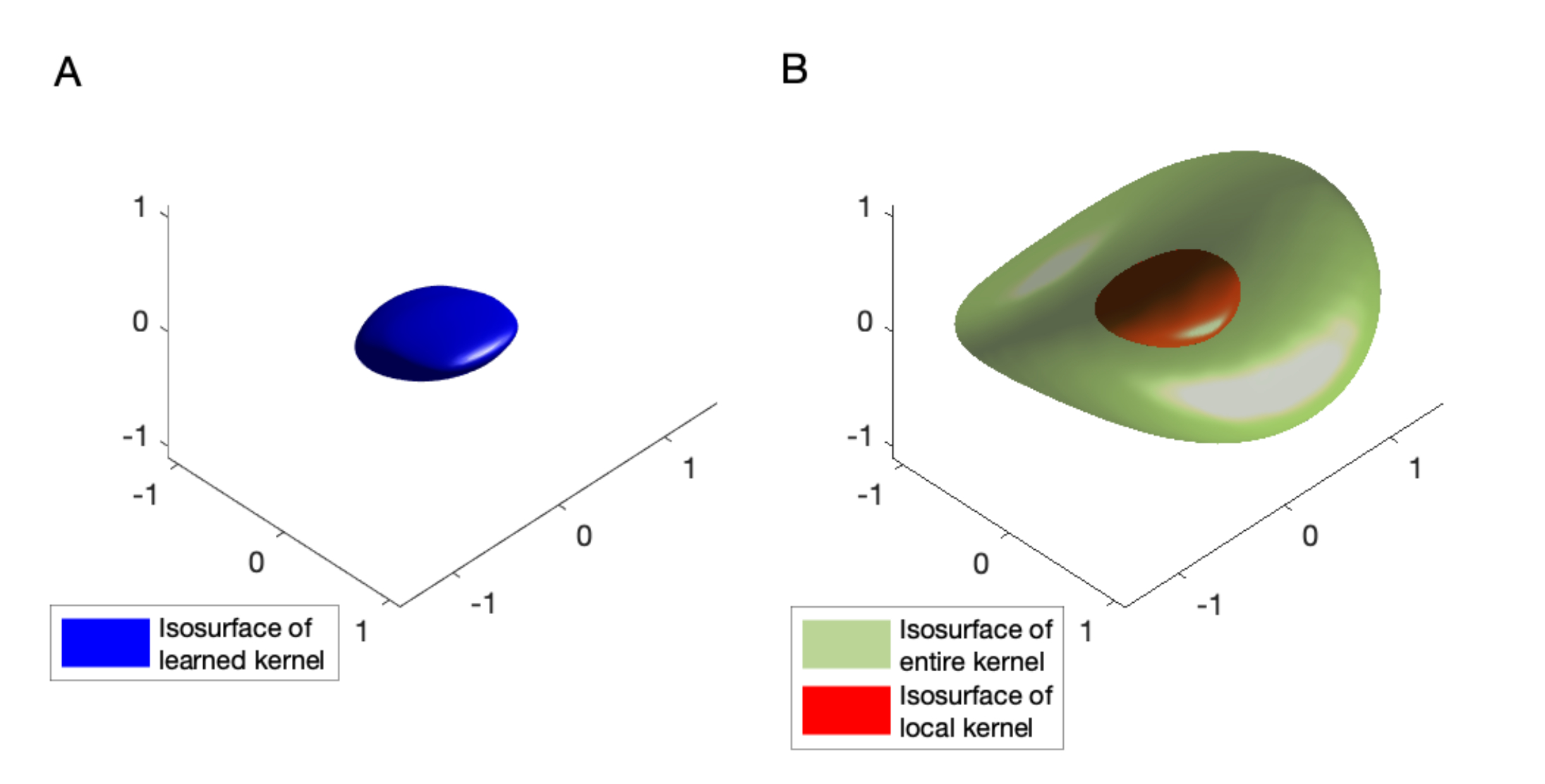}
    \caption{ \textbf{(A)} Isosurface of the transition kernel $K^1$. 
    \textbf{(B)} Isosurfaces of heat equation solution. In green: isosurface of the entire solution. In red: isosurface of the local solution.
    As expected  the transition kernel is closer to this local isosurface.}
  \label{fig:Isos_kernel}
\end{figure}

The 3-D counterpart of this visualization is shown in Figure \ref{fig:Isos_kernel}A, displaying two isosurfaces of the sub-Riemannian kernel, whereas the Figure \ref{fig:Isos_kernel}B shows an isosurface of $K^1$. The geometry of the learned kernel $K^1$ is qualitatively similar to the local behavior of the theoretical kernel, shown by the red isosurface in Figure \ref{fig:Isos_kernel}A.

\section{Discussion } \label{Sym_Discussion}           

In this work, we showed how approximate group invariances arise in the early layers of a biologically inspired CNN architecture during learning on natural images, and we established a parallel with the architecture and plasticity of the early visual pathway. First, the LoG-shaped filter emerging in the pre-filtering stage $\ell^0$ closely resembles the typical LGN receptive profile exhibiting a rotational symmetry. The presence of $\ell^0$ also enhanced the orientation tuning of first-layer filters, thus separating the first image elaboration into two steps that may be roughly associated with sub-cortical and early cortical processing. Moreover, the introduction of lateral connections in the first network layer allowed us to investigate the relationship between the geometry of the feedforward filters and the selectivity of the learned connectivity kernel $K^1$. Indeed, the Gabor approximation of first layer filters provided us with a set of coordinates to re-map the kernel $K^1$ into the $\mathbb{R}^2\times S^1$ feature space, thus allowing to express the connectivity strength in terms of relative positions and orientations of the filters. Strikingly, the association fields induced by the learned connectivity bear a close resemblance to those obtained from edge co-occurrence in natural images by \cite{edge-stat}, as well as to the flow of neural activity described by a sub-Riemannian heat equation in the Lie group setting introduced by \cite{cs06}.

We stress that such geometric properties arose spontaneously during the training of the CNN architecture on a dataset of natural images, the only constraint being the translation-invariance imposed by the convolutional structure of the network.

As future perspectives, we would like to extend our study by considering a wider range of features -- possibly leading to a finer characterization of the geometry of first layer filters, including more complex types of receptive profiles. Moreover, considering a larger images dataset would make it possible to examine horizontal connectivity patterns spread over a wider spatial region. Finally, another direction could be to extend our model by comparing the properties of deeper network layers to the processing in higher cortices of the visual system.

\bibliographystyle{apalike}
\bibliography{Sym_refs}{}

\end{document}